\documentclass[copyright,creativecommons]{eptcs}
\providecommand{\event}{QAPL 2016} 

\usepackage{amsmath}
\usepackage{amssymb}
\usepackage{amsthm}
\usepackage{verbatim} 
\usepackage{stmaryrd}
\usepackage{paralist}
\usepackage{cite}
\usepackage{appendix}
\usepackage{url}
\usepackage[]{units}
\usepackage{tikz}
\usepackage{subfig}
\usepackage{array}
\usepackage{color}
\usepackage{empheq}
\usetikzlibrary{calc,arrows,positioning}
\allowdisplaybreaks

\newif\ifdraft\drafttrue

\DeclareFontFamily{U}{mathb}{\hyphenchar\font45}
\DeclareFontShape{U}{mathb}{m}{n}{
      <5> <6> <7> <8> <9> <10> gen * mathb
      <10.95> mathb10 <12> <14.4> <17.28> <20.74> <24.88> mathb12
}{}
\DeclareSymbolFont{mathb}{U}{mathb}{m}{n}
\DeclareMathSymbol{\sqdoublecup} {2}{mathb}{"5F} 
\DeclareMathSymbol{\boxplus} {2}{mathb}{"60} 

\DeclareMathOperator{\logic}{\mathcal L}

\newcommand{\sgotopi}{(s, a, \pi) \in \rightarrow}
\newcommand{\snotdob}{b \not\in \init{s}}

\newcommand{\E}{\mathcal E}

\newcommand{\w}{\mathfrak{w}}
\newcommand{\W}{\mathfrak{W}}

\newcommand{\diam}[1]{\langle #1 \rangle}

\DeclareMathOperator{\dist}{\mathbf{d}_{\lambda}}
\DeclareMathOperator{\distfs}{\mathfrak{d}_{\lambda}}
\DeclareMathOperator{\distfd}{\mathfrak{D}_{\lambda}}
\newcommand{\distk}[1]{\mathbf{d}_{\lambda}^{#1}}
\newcommand{\distfsk}[1]{\mathfrak{d}_{\lambda}^{#1}}
\newcommand{\distfdk}[1]{\mathfrak{D}_{\lambda}^{#1}}

\newcommand{\depth}[1]{\mathrm{dpt}(#1)}
\newcommand{\init}[1]{\mathrm{init}(#1)}
\newcommand{\der}[1]{\mathrm{der}(#1)}

\newcommand{\rel}{\,{\mathcal R}\,}
\newcommand{\reldist}{\,{\mathcal R}^{\dagger}\,}

\newcommand{\proc}{\mathbf{S}}
\newcommand{\ProbDist}[1]{\Delta(#1)}

\newcommand{\Act}{\mathcal A}

\newcommand{\eqlog}{\,\equiv_{\logic}\,}

\newcommand{\trans}[1][]{\xrightarrow{\, {#1} \, }}
\newcommand{\ntrans}[1][]{\mathrel{{\trans[#1]}\makebox[0em][r]{$\not$\hspace{2ex}}}{\!}}

\newcommand{\bisim}{\sim}

\newcommand{\N}{\mathbb{N}} 

 \DeclareMathOperator{\Kantorovich}{\mathbf{K}}
 \DeclareMathOperator{\Hausdorff}{\mathbf{H}}
 \DeclareMathOperator{\Bisimulation}{\mathbf{B}}
\DeclareMathOperator{\logicstate}{\logic^{\mathrm{s}}}
\DeclareMathOperator{\logicdist}{\logic^{\mathrm{d}}}

\newcommand{\leaveout}[1]{}

\newcommand{\Red}[2]{\textrm{Red}}

\newcommand{\support}{\mathsf{supp}}

\newcommand{\topd}{\ensuremath{\mathbf{1}}}
\newcommand{\botd}{\ensuremath{\mathbf{0}}}

\newtheorem{theorem}{Theorem}
\newtheorem{proposition}{Proposition}
\newtheorem{lemma}{Lemma}
\newtheorem{corollary}{Corollary}

\theoremstyle{definition}
\newtheorem{definition}{Definition}
\newtheorem{example}{\emph{Example}}

\theoremstyle{remark}

\title{Logical Characterization of Bisimulation Metrics}
\author{Valentina Castiglioni
\institute{University of Insubria (IT)}
\email{v.castiglioni2@uninsubria.it}
\and
Daniel Gebler
\institute{VU University Amsterdam (NL)}
\email{e.d.gebler@vu.nl}
\and
Simone Tini
\institute{University of Insubria (IT)}
\email{simone.tini@uninsubria.it}
}

\def\titlerunning{Logical Characterization of Bisimulation Metrics}
\def\authorrunning{V. Castiglioni, D. Gebler \& S. Tini}

\begin{document}

\maketitle

\begin{abstract}
Bisimulation metrics provide a robust and accurate approach to study the behavior of nondeterministic probabilistic processes.
In this paper, we propose a logical characterization of bisimulation metrics based on a simple probabilistic variant of the Hennessy-Milner logic.
Our approach is based on the novel notions of mimicking formulae and distance between formulae.
The former are a weak version of the well known characteristic formulae and allow us to characterize also (ready) probabilistic simulation and probabilistic bisimilarity.
The latter is a $1$-bounded pseudometric on formulae that mirrors the Hausdorff and Kantorovich lifting the defining bisimilarity pseudometric.
We show that the distance between two processes equals the distance between their own mimicking formulae.
\end{abstract}


\section{Introduction}

To verify whether the implementation proposed for a reactive concurrent system (or process) conforms the properties and behavior entailed by its specification, two main approaches have been proposed: \emph{equivalence checking}\cite{AILS12} and \emph{model checking}\cite{CE81}.
The former is used when both the implementation and specification of the system are modeled by a labeled transition system (LTS)\cite{K76} and it consists in verifying whether each of the observations enabled by the implementation is allowed by the specification. 
To this aim, several notions of behavioral equivalence, identifying those processes which are indistinguishable to external observers, and of behavioral preorder have been proposed.
Among those equivalences, the most successfully employed is the bisimulation equivalence \cite{P81}.
In model checking the specification of the system is expressed as a formula in a modal or temporal logic \cite{E90}, whereas the actual behavior of the system is still defined by an LTS.
Thus, the LTS meets its specification if it is a model for the formula.

Logical characterizations of equivalences constitute the connection between these two approaches and consist in showing that the process equivalence induced by a logic matches the behavioral equivalence.
Given any equivalence, a logic is said to be \emph{adequate} for a behavioral equivalence if two processes are behaviorally equivalent if and only if they satisfy the same formulae. 
Moreover, a logic is said to be \emph{expressive} for a behavioral equivalence if for each process $s$ we can build a characteristic formula $\phi_s$ for the considered equivalence \cite{GS86a} such that any process $t$ is equivalent to $s$ if and only if $t$ satisfies $\phi_s$.
The first example of a logical characterization is given by the class of modal formulae known as Hennessy-Milner logic (HML)\cite{HM85}, which is proved to be adequate for bisimulation equivalence.

In this paper we study logical characterizations in the field of probabilistic concurrent systems.
As semantic model, we consider nondeterministic probabilistic labeled transition systems (PTSs) \cite{S95}, which extend LTSs to allow us to model the behavior of those systems in which nondeterminism and probability coexist.
As a notion of behavioral equivalence we consider probabilistic bisimulation (see, e.g., \cite{LS91,SL95, MCL12}, and \cite{H12} for a survey).
In the literature there are several examples of logics that are adequate for probabilistic bisimulation and that consider different semantic models: reactive probabilistic transition systems in the seminal work \cite{LS91}, probabilistic automata in \cite{PS07,HPSWZ11}, PTSs in \cite{BdNL15,DD11,DvG10}, labeled Markov processes (LMP) in \cite{DEP02,dATW12} and continuous-time continuous-space LMP in \cite{MCL12}.
An expressive characterization of probabilistic bisimulation in the context of PTSs is given in \cite{DvG10} where the construction of characteristic formulae is allowed by a very rich modal logic, the probabilistic modal $\mu$-calculus.

Probabilistic bisimulations have been proposed as simple and elegant relations proving whether the behavior of two probabilistic systems is  exactly the same.
However, probabilistic systems exhibit quantitative properties and one may be interested in quantifying how much the current implementation of a system is far from its intended specification.
This is achieved through the so called \emph{behavioral pseudometrics} \cite{BW01b,DGJP04,DJGP02,GJS90,KN96}, which have been showed to provide a robust semantics for probabilistic processes \cite{BW05,DCPP06,DEP02,DGJP04,GT14,GLT15,GLT16,GT15}.
In particular, bisimulation pseudometrics (usually simply denoted as bisimulation metrics) are 1-bounded pseudometrics that are the quantitative analogue to probabilistic bisimulation and assign to each pair of processes a distance which measures the proximity of their quantitative properties and whose kernel is probabilistic bisimulation.
In \cite{GLT15,GT15,GLT16} it is proved that bisimulation metric is also suitable for compositional reasoning and thus for the specification and verification of probabilistic systems.
Logical characterizations for bisimulation metrics have been proposed in \cite{BW05,AFS04,BBLM14} based on real-valued logics and in \cite{LMP12} based on a boolean-valued logic together with a notion of distance between formulae defined in terms of a given ground distance between processes. 

In this paper we consider the boolean-valued logic $\logic$ of \cite{DD11}, and by means of a novel notion of \emph{mimicking formula}, we show that $\logic$ allows us to characterize bisimulation metrics and to provide also logical characterizations of probabilistic bisimulation, ready simulation and simulation.
The notion of mimicking formula is given for each process as a formula in $\logic$ that captures the ability and the inability of the process to execute any action and describes the probabilistic behavior of the process.
Then we introduce a notion of \emph{distance between formulae} in $\logic$, which is a $1$-bounded pseudometric assigning to each pair of formulae a suitable quantitative analogue of their syntactic disparities. 
This immediately gives us a notion of distance between processes, which we call \emph{logical distance} and corresponds to the distance between the respective mimicking formulae.
Our main result is that the logical distance between two processes corresponds to their distance as expressed by the bisimulation metric.
Up to our knowledge, this is the first characterization of bisimulation metric given by means of a boolean-valued logic and of a distance on the logic not defined in terms of the distance between processes.
Then, we also show that the logic $\logic$ is \emph{weak expressive} for probabilistic bisimulation, meaning that two processes are probabilistic bisimilar if and only if their mimicking formulae are equivalent under a proper definition of structural equivalence over $\logic$.
We cannot refer to this characterization of probabilistic bisimulation as to an expressive one since the single mimicking formula of a process is not powerful enough to capture the whole equivalence class of the process, namely it is not its characteristic formula for probabilistic bisimulation.
We remark that the fully expressive characterization of bisimulation of \cite{DvG10} requires a logic much richer that $\logic$. 
Finally, we show that the logic $\logic$ is expressive for probabilistic ready simulation, meaning that the mimicking formula of a process is its characteristic formula for probabilistic ready simulation, and that $\logic$ is expressive for simulation, meaning that the negation-free version of the mimicking formula of a process is its characteristic formula for probabilistic simulation. 
Up to our knowledge, this is the first paper where a single logic is used to characterize both bisimulation metric and classic notions of equivalence and preorders.

Summarizing, by means of mimicking formulae for processes defined with the logic $\logic$ we get:
\begin{enumerate}
\item Weak expressive characterization of probabilistic bisimulation: we establish whether two processes are probabilistic bisimilar by simply comparing their mimicking formulae (Theorem~\ref{thm:strong_characterization}).
\item Expressive characterization of ready probabilistic simulation (Theorem~\ref{thm:ready_sim_char}) and probabilistic simulation (Theorem~\ref{thm:characteristic_sim}).
\item Characterization of bisimulation metric: we define a distance between formulae and we prove that the bisimulation distance between two processes equals the distance between their mimicking formulae (Theorem~\ref{thm:ell=bisim}).
\end{enumerate}

The paper is organized as follows.
Section~\ref{sec:background} contains the background.
In Section~\ref{sec:logic} we introduce the logic $\logic$ of \cite{DD11}.
In Section~\ref{sec:bis_char} we present mimicking formulae and the characterization of probabilistic bisimulation and (ready) simulation. 
In Section~\ref{sec:metric_char} we introduce the distance on $\logic$ and our main result, i.e. the modal characterization of bismulation metric.
In Section \ref{sec:conclusion} we discuss related and future work.


\section{Background}
\label{sec:background}

Nondeterministic probabilistic transition systems \cite{S95} combine labeled transition systems and discrete time Markov chains, thus allowing us to model separately the reactive behavior, nondeterministic choices and probabilistic choices.

As state space we take a set $\proc$, whose elements are called $\emph{processes}$.
We let $s,t,\ldots$ range over $\proc$.  
Probability distributions over $\proc$ are mappings $\pi \colon \proc \to [0,1]$ with $\sum_{s \in \proc} \pi(s) = 1$ that assign to each $s \in \proc$ its probability $\pi(s)$. 
By $\ProbDist{\proc}$ we denote the set of all probability distributions over $\proc$.
We let $\pi, \pi',\dots$ range over $\ProbDist{\proc}$.
For $\pi \in \ProbDist{\proc}$, we denote by $\support(\pi)$ the support of $\pi$, namely $\support(\pi) = \{ s \in \proc \mid \pi(s) >0\}$. 
We consider only probability distributions with \emph{finite} support.
For $s \in \proc$ we denote by $\delta_s$ the \emph{Dirac distribution} defined by $\delta_s(s)= 1$ and $\delta_s(t)=0$ for $s \neq t$.
The convex combination $\sum_{i \in I} p_i \pi_i$ of a family $\{\pi_i\}_{i \in I}$ of probability distributions $\pi_i \in \ProbDist{\proc}$ with $p_i \in (0,1]$ and $\sum_{i \in I} p_i = 1$ is defined by $(\sum_{i \in I} p_i \pi_i)(s) = \sum_{i \in I} (p_i \pi_i(s))$ for all $s \in \proc$. 

\begin{definition}[PTS, \cite{S95}]
A \emph{nondeterministic probabilistic labeled transition system (PTS)} is a triple $(\proc,\Act,\trans[])$, where: 
\begin{inparaenum}[(i)]
\item $\proc$ is a countable set of processes, 
\item $\Act$ is a countable set of \emph{actions}, and 
\item $\trans[] \subseteq {\proc \times \Act \times \ProbDist{\proc}}$ is a \emph{transition relation}. 
\end{inparaenum}
\end{definition}

We call $(s,a,\pi)\in\trans[]$ a \emph{transition}, and we write $s\trans[a]\pi$ for $(s,a,\pi) \in\trans[]$.
We write $s \trans[a] $ if there is a distribution $\pi \in \ProbDist{\proc}$ with $s \trans[a] \pi$, and $s \ntrans[a]$ otherwise. 
Let $\init{s} =\{ a \in \Act \mid  s\trans[a]\}$ denote the set of the actions that can be performed by $s$.
Let $\der{s,a} =\{\pi\in\ProbDist{\proc} \mid s\trans[a]\pi\}$ denote the set of the distributions reachable from $s$ through action $a$.
Let $\depth{s}$ denote the \emph{depth} of $s$, namely the maximal number of sequenced transitions that can be performed from $s$, defined by $\depth{s} = 0$, if $\init{s} = \emptyset$, and $\depth{s} = 1+ \sup_{a \in \init{s}, \pi \in \der{s,a}, t \in \support(\pi)} \depth{t}$, otherwise.
We say that a process $s \in \proc$ is \emph{image-finite} if for all actions $a \in\init{s}$ the set $\der{s,a}$ is finite \cite{HPSWZ11}, and that $s$ has \emph{finite depth} if $\depth{s}$ is finite.
Finally, we denote as \emph{finite} the image-finite processes with finite depth. 


\subsection{Probabilistic (bi)simulations}
A probabilistic bisimulation is an equivalence relation over $\proc$ that equates processes $s,t \in \proc$ if they can mimic each other's transitions and evolve to distributions that are in turn related by the same bisimulation.
To formalize this, we need to lift relations over processes to relations over distributions.
\begin{definition}[Relation lifting, \cite{DvG10}]\label{def:relation_lifting}
\label{def:relational_lifting}
The \emph{lifting} of a relation $\rel \subseteq \proc \times \proc$ is the relation $\reldist \subseteq \ProbDist{\proc} \times \ProbDist{\proc}$ with $\pi \reldist \pi'$ whenever there is 
a set of indexes $I$ such that 
\[ 
\text{(i) } \pi = \sum_{i \in I} p_i \delta_{s_i}\text{, \quad (ii) } \pi' = \sum_{i \in I} p_i  \delta_{t_i}, \; \; \text{  and \quad (iii) } s_i \; \rel \; t_i \mbox{ for all } i \in I\text{.}
\]
\end{definition}

\begin{definition}
[Probabilistic (bi)simulations, \cite{S95,LS91}]
\label{def:prob_(bi)sim}
Assume a PTS.
\begin{enumerate}
\item
A binary relation $\rel \,\subseteq \proc \times \proc$ is a \emph{probabilistic simulation} if whenever $s \rel  t$
\begin{center} 
if $s \trans[a] \pi_s$ then there is a transition $t \trans[a] \pi_t$ such that  
$\pi_s \rel^{\dagger} \pi_t$.
\end{center}
\item
A probabilistic simulation $\rel$ is a \emph{probabilistic ready simulation} if whenever $s \rel t$
\begin{center} 
if $s \ntrans[a]$ then $t \ntrans[a]$.
\end{center}
\item
A \emph{probabilistic bisimulation} is a symmetric probabilistic simulation.
\end{enumerate}
\end{definition}
The union of all probabilistic simulations (resp.: ready simulations, bisimulations) is the greatest probabilistic simulation (resp.: ready simulation, bisimulation),  denoted  $\sqsubseteq$ (resp.: $\sqsubseteq^r$, $\bisim$), called \emph{similarity} (resp.: \emph{ready similarity}, \emph{bisimilarity}), and is a preorder (resp.: preorder, equivalence).

\begin{example}
\label{ex:bisim}
Consider the processes $s,t \in \proc$ represented in Figure~\ref{fig:ex_bisim}.
We have that $s \sim t$.
\begin{figure}
\begin{center}
\begin{tikzpicture}
\node at (0,7.3){$\boldsymbol s$};
\draw[->](0,7.1)--(0,6.5);
\node at (0.2,6.9){$\boldsymbol a$};
\node at (5,7.3){$\boldsymbol t$};
\draw[->](5,7.1)--(3.5,6.5);
\node at (4.2,6.9){$\boldsymbol a$};
\draw[->](5,7.1)--(6.5,6.5);
\node at (5.85,6.9){$\boldsymbol a$};
\draw[dotted, thick, ->](0,6.5)--(0,5.7);
\node at (0.2,6.2){$\boldsymbol 1$};
\draw[dotted, thick, ->](3.5,6.5)--(2.5,5.7);
\node at (2.85,6.3){$\boldsymbol{\frac{3}{4}}$};
\draw[dotted, thick, ->](3.5,6.5)--(4.5,5.7);
\node at (4.25,6.3){$\boldsymbol{\frac{1}{4}}$};
\draw[dotted, thick, ->](6.5,6.5)--(6.5,5.7);
\node at (6.7,6.2){$\boldsymbol 1$};
\node at (0,5.4){$\boldsymbol{s_1}$};
\draw[->](0,5.2)--(0,4.4);
\node at (0.2,4.85){$\boldsymbol{b}$};
\node at (2.5,5.4){$\boldsymbol{t_1}$};
\draw[->](2.5,5.2)--(1.5,4.4);
\node at (1.8,4.85){$\boldsymbol{b}$};
\draw[->](2.5,5.2)--(3.5,4.4);
\node at (3.2,4.85){$\boldsymbol{b}$};
\node at (4.5,5.4){$\boldsymbol{t_2}$};
\draw[->](4.5,5.2)--(4.5,4.4);
\node at (4.7,4.85){$\boldsymbol{b}$};
\node at (6.5,5.4){$\boldsymbol{t_3}$};
\draw[->](6.5,5.2)--(6.5,4.4);
\node at (6.7,4.85){$\boldsymbol{b}$};
\draw[dotted, thick,->] (0,4.4)--(0,3.7);
\node at (0.2,4.1){$\boldsymbol{1}$};
\node at (0,3.5){$\boldsymbol{\mathrm{nil}}$};
\draw[dotted, thick, ->](1.5,4.4)--(1.5,3.7);
\node at (1.7,4.1){$\boldsymbol{1}$};
\node at (1.5,3.5){$\boldsymbol{\mathrm{nil}}$};
\draw[dotted, thick, ->](3.5,4.4)--(3.5,3.7);
\node at (3.7,4.1){$\boldsymbol{1}$};
\node at (3.5,3.5){$\boldsymbol{\mathrm{nil}}$};
\draw[dotted, thick, ->](4.5,4.4)--(4.5,3.7);
\node at (4.7,4.1){$\boldsymbol{1}$};
\node at (4.5,3.5){$\boldsymbol{\mathrm{nil}}$};
\draw[dotted, thick,->] (6.5,4.4)--(6.5,3.7);
\node at (6.7,4.1){$\boldsymbol{1}$};
\node at (6.5,3.5){$\boldsymbol{\mathrm{nil}}$};
\end{tikzpicture}
\end{center}
\caption{\label{fig:ex_bisim} Processes $s,t \in \proc$ are probabilistic bisimilar.}
\end{figure}
It is immediate to verify that processes $t_1,t_2,t_3$ are all bisimilar to process $s_1$, since they can execute only $b$-labeled transitions reaching with probability $1$ the process $\mathrm{nil}$, namely the process which cannot execute any action.
As a consequence, we can directly conclude that $\delta_{s_1} \sim^{\dagger} \delta_{t_3}$.
Likewise, it is quite easy to see that the Dirac distribution on $s_1$ can be rewritten as the convex combination $\delta_{s_1} = \frac{3}{4} \delta_{s_1} + \frac{1}{4} \delta_{s_1}$.
Hence if we let $\pi = \frac{3}{4} \delta_{t_1} + \frac{1}{4} \delta_{t_2}$ be the probability distribution to which process $t$ evolves by executing the leftmost $a$-labeled transition, from $s_1 \sim t_1$ and $s_1 \sim t_2$ we can conclude that $\delta_{s_1} \sim^{\dagger} \pi$ and thus $s \sim t$.
\qed
\end{example}

These equivalences and preorders are approximated by relations that consider only the first $k$ transition steps \cite{B98,HPSWZ11}.

\begin{definition}
[Up-to-$k$ (bi)simulations]
\label{def:up-to-k-bis}
Assume a PTS.
\begin{enumerate}
\item
The family of the \emph{up-to-$k$ simulations} $\sqsubseteq_k$, for $k \in \N$, is inductively defined as follows:
\begin{enumerate}
\item $\sqsubseteq_0 = \proc \times \proc$;
\item $s \sqsubseteq_{k+1} t$ if whenever $s \trans[a] \pi_s$ there is a transition $t \trans[a] \pi_t$ such that 
$\pi_s \sqsubseteq_k^{\dagger} \pi_t$.
\end{enumerate}
\item
The family of the \emph{up-to-$k$ ready simulations} $\sqsubseteq_k^r$, for $k \in \N$, is inductively defined as follows:
\begin{enumerate}
\item $\sqsubseteq_0^r = \proc \times \proc$;
\item $s \sqsubseteq_{k+1}^r t$ if whenever $s \trans[a] \pi_s$ there is a transition $t \trans[a] \pi_t$ such that 
$\pi_s {\sqsubseteq_k^{r}}^{\dagger} \pi_t$, and whenever $s \ntrans[a]$ then $t \ntrans[a]$.
\end{enumerate}
\item
The \emph{up-to-$k$ bisimulation} $\sim_k$ is the kernel of $\sqsubseteq_k$.
\end{enumerate}
Finally, we define $\sqsubseteq_{\omega} = \bigcap_{k \ge 0} \sqsubseteq_k$, $\sqsubseteq^r_{\omega} = \bigcap_{k \ge 0} \sqsubseteq^r_k$, and $\sim_{\omega} = \bigcap_{k \ge 0} \sim_k$.
\end{definition}

\begin{proposition}[\!\protect{\cite{HPSWZ11}}]
\label{prop:bis_lim}
On image-finite PTSs, $\sqsubseteq_{\omega}$ (resp.: $\sqsubseteq^r_{\omega}$, $\sim_{\omega}$), coincides with $\sqsubseteq$ (resp.: $\sqsubseteq^r$, $\sim$).
\end{proposition}


\subsection{Bisimulation metrics}
Probabilistic bisimulations answer the question of whether two processes behave precisely the same way or not.
Bisimulation metrics answer the more general question of how far the behavior of two processes is apart.
They are defined as $1$-bounded pseudometrics on $\proc$ giving the distance between processes.
\begin{definition}
[1-bounded pseudometric]
\label{def:pseudometric}
A function $d \colon \proc \times \proc \to [0,1]$ is a \emph{1-bounded pseudometric} if 
\begin{inparaenum}[(i)]
\item $d(s,s) =0$,
\item $d(s,t) = d(t,s)$, and 
\item $d(s,t) \le d(s,p) + d(p,t)$ for all $s,t,p \in \proc$.
\end{inparaenum}
\end{definition}

Bisimilarity metric is defined as the least fixed point of a suitable functional on the following structure.
Let ${([0,1]^{\proc \times \proc},\sqsubseteq)}$ be the complete lattice of functions 
$d \colon \proc \times \proc \to [0,1]$
ordered by $d_1 \sqsubseteq d_2$ iff $d_1(s,t) \le d_2(s,t)$ for all processes $s,t \in \proc$. 
Then for each set $D \subseteq [0,1]^{\proc \times \proc}$ the supremum and infinimum are 
$\sup(D)(s,t) = \sup_{d \in D}d(s,t)$ and 
$\inf(D)(s,t) = \inf_{d \in D}d(s,t)$ for all $s,t \in \proc$.
The bottom element is the function $\botd$ with $\botd(s,t)=0$, and the top element is the function $\topd$ with $\topd(s,t)=1$, for all $s,t \in \proc$.

Bisimulation metrics are characterized using the quantitative analogous of the bisimulation game, meaning that two process $s,t  \in \proc$ at some given distance can mimic each other's transitions and evolve into distributions that are at distance not greater than the distance between $s$ and $t$. 
To formalize this, we need a notion that lifts pseudometrics from processes to distributions.

A \emph{matching} for distributions $\pi,\pi' \in \ProbDist{\proc}$ is a distribution over the product state space $\w \in \Delta(\proc \times \proc)$ with $\pi$ and $\pi'$ as left and right marginal, namely $\sum_{t\in \proc} \w(s,t)=\pi(s)$ and $\sum_{s\in \proc} \w(s,t)=\pi'(t)$ for all $s,t \in \proc$. 
Let $\W(\pi,\pi')$ denote the set of all matchings for $\pi,\pi'$.
Intuitively, a matching $\w \in \W(\pi,\pi')$ may be understood as a transportation schedule describing the shipment of probability mass from $\pi$ to $\pi'$. This motivation dates back to the Monge-Kantorovich optimal transport problem~\cite{Vil08}.

\begin{definition}
[Kantorovich lifting, \cite{K42}]
\label{def:Kantorovich}
Let $d\colon \proc \times \proc \to [0,1]$ be a 1-bounded pseudometric. 
The \emph{Kantorovich lifting} of $d$ is the 1-bounded pseudometric $\Kantorovich(d)\colon \ProbDist{\proc} \times \ProbDist{\proc} \to [0,1]$ defined by
\[
\Kantorovich(d)(\pi,\pi') = \min_{\w \in \W(\pi,\pi')} \sum_{s,t\in \proc}\w(s,t)\cdot d(s,t) 
\]
for all $\pi,\pi' \in \ProbDist{\proc}$.
For any $1$-bounded pseudoemetric $d$, we call $\Kantorovich(d)$ the \emph{Kantorovich pseudometric}.
\end{definition}
We remark that accordingly to the original definition, we should have defined the Kantorovich pseudometric as $\Kantorovich(d)(\pi,\pi') = \inf_{\w \in \W(\pi,\pi')} \sum_{s,t\in \proc}\w(s,t)\cdot d(s,t)$, namely as th infimum over the matchings for $\pi$ and $\pi'$, for any $\pi,\pi' \in \ProbDist{\proc}$ and $1$-bounded pseudometric $d$. 
However, the assumption of having only probability distributions with a finite support guarantees that the infimum is always achieved, since there can be only finitely many matchings between the two distributions, and therefore it is a minimum.
As a consequence, the continuity of the lifting functional $\Kantorovich$ is guaranteed \cite{vB12}.

In order to capture nondeterministic choices, we need to lift pseudometrics on distributions to pseudometrics on sets of distributions.

\begin{definition}
[Hausdorff lifting]
\label{def:Hausdorff}
Let $\hat{d} \colon \ProbDist{\proc} \times \ProbDist{\proc} \to [0,1]$ be a 1-bounded pseudometric.
The \emph{Hausdorff lifting} of $\hat{d}$ is the $1$-bounded pseudometric $\Hausdorff(\hat{d})\colon P(\ProbDist{\proc}) \times P(\ProbDist{\proc}) \to [0,1]$ defined by
\[
\Hausdorff(\hat{d})(\Pi_1,\Pi_2) = \max \Big\{ \adjustlimits\sup_{\pi_1 \in \Pi_1}\inf_{\pi_2 \in \Pi_2} \hat{d}(\pi_1,\pi_2), \adjustlimits\sup_{\pi_2\in \Pi_2}\inf_{\pi_1\in \Pi_1} \hat{d}(\pi_2,\pi_1) \Big\}
\]
for all $\Pi_1,\Pi_2 \subseteq \ProbDist{\proc}$, where $\inf \emptyset = 1$, $\sup \emptyset = 0$.
For any $1$-bounded pseudometric $\hat{d}$, we call $\Hausdorff(\hat{d})$ the \emph{Hausdorff pseudometric}.
\end{definition}

The quantitative analogue of the bisimulation game is defined by means of a functional $\Bisimulation$ over the lattice ${([0,1]^{\proc \times \proc},\sqsubseteq)}$.
By means of a \emph{discount factor} $\lambda \in (0,1]$, $\Bisimulation$ allows us to specify how much the behavioral distance of future transitions is taken into account to determine the distance between two processes~\cite{AHM03,DGJP04}. 
The discount factor $\lambda=1$ expresses no discount, meaning that the differences in the behavior between $s$ and $t$ are considered irrespective of after how many steps they can be observed.

\begin{definition}[Bisimulation metric functional] \label{def:metric_bisim_functional}
Let $\Bisimulation \colon [0,1]^{\proc \times \proc} \to [0,1]^{\proc \times \proc}$ be the function defined by
\[
	\Bisimulation(d)(s,t) = \sup_{a\in A} \left\{ \Hausdorff(\lambda \cdot \Kantorovich(d))(\der{s,a}, \der{t,a}) \right\}
\]
for $d \colon \proc \times \proc \to [0,1]$ and $s,t \in \proc$, with
$(\lambda \cdot \Kantorovich(d))(\pi,\pi')=\lambda \cdot \Kantorovich(d)(\pi,\pi')$.
\end{definition}
We remark that since the sets $\der{s,a}$ and $\der{t,a}$ are finite for all $a \in \Act$, $s,t \in \proc$, due to the image-finiteness assumption, the suprema and infima in the definition of the Hausdroff pseudometric are always achieved, thus becoming maxima and minima, respectively.
Hence, considering that the lifting functional $\Kantorovich$ is continuous, the continuity of the lifting functional $\Hausdorff$ is guaranteed \cite{vB12}.

It is not hard to show that $\Bisimulation$ is monotone. 
Then, since $([0,1]^{\proc \times \proc},\sqsubseteq)$ is a complete lattice, by the Knaster-Tarski theorem $\Bisimulation$ has the least fixed point.
Bisimulation metrics are the 1-bounded pseudometrics being prefixed points of $\Bisimulation$ \cite{DCPP06}.
The bisimilarity metric is defined as the least fixed point of $\Bisimulation$, and is a 1-bounded pseudometric \cite{DCPP06}.
Hence, bisimilarity metric is the least bisimulation metric.

\begin{definition}[Bisimulation metric, \cite{DCPP06}] \label{def:bisimilarity_metric}
A 1-bounded pseudometric $d \colon \proc \times \proc \to [0,1]$ is a \emph{bisimulation metric} if{f} $\Bisimulation(d) \sqsubseteq d$.
The least fixed point of $\Bisimulation$ is denoted by $\dist$ and called the \emph{bisimilarity metric}.
\end{definition}

\begin{example}
\label{ex:dist}
Consider the process $s \in \proc$ from previous Example~\ref{ex:bisim} and the process $s' \in \proc$, both represented in Figure~\ref{fig:ex_dist}.
\begin{figure}
\begin{center}
\begin{tikzpicture}
\node at (0,7.3){$\boldsymbol s$};
\draw[->](0,7.1)--(0,6.5);
\node at (0.2,6.9){$\boldsymbol a$};
\node at (5,7.3){$\boldsymbol s'$};
\draw[->](5,7.1)--(3.5,6.5);
\node at (4.2,6.9){$\boldsymbol a$};
\draw[->](5,7.1)--(6.5,6.5);
\node at (5.85,6.9){$\boldsymbol a$};
\draw[dotted, thick, ->](0,6.5)--(0,5.7);
\node at (0.2,6.2){$\boldsymbol 1$};
\draw[dotted, thick, ->](3.5,6.5)--(2.5,5.7);
\node at (2.85,6.3){$\boldsymbol{\frac{3}{4}}$};
\draw[dotted, thick, ->](3.5,6.5)--(4.5,5.7);
\node at (4.25,6.3){$\boldsymbol{\frac{1}{4}}$};
\draw[dotted, thick, ->](6.5,6.5)--(5.5,5.7);
\node at (5.8,6.3){$\boldsymbol{\frac{1}{2}}$};
\draw[dotted, thick,->](6.5,6.5)--(7.5,5.7);
\node at (7.2,6.3){$\boldsymbol{\frac{1}{2}}$};
\node at (0,5.4){$\boldsymbol{s_1}$};
\draw[->](0,5.2)--(0,4.4);
\node at (0.2,4.85){$\boldsymbol{b}$};
\node at (2.5,5.4){$\boldsymbol{s_2}$};
\draw[->](2.5,5.2)--(1.5,4.4);
\node at (1.8,4.85){$\boldsymbol{b}$};
\draw[->](2.5,5.2)--(3.5,4.4);
\node at (3.2,4.85){$\boldsymbol{c}$};
\node at (4.5,5.4){$\boldsymbol{s_3}$};
\draw[->](4.5,5.2)--(4.5,4.4);
\node at (4.7,4.85){$\boldsymbol{b}$};
\node at (5.5,5.4){$\boldsymbol{s_4}$};
\draw[->](5.5,5.2)--(5.5,4.4);
\node at (5.7,4.85){$\boldsymbol{b}$};
\node at (7.5,5.4){$\boldsymbol{s_5}$};
\draw[->](7.5,5.2)--(7.5,4.4);
\node at (7.7,4.85){$\boldsymbol{c}$};
\draw[dotted, thick,->] (0,4.4)--(0,3.7);
\node at (0.2,4.1){$\boldsymbol{1}$};
\node at (0,3.5){$\boldsymbol{\mathrm{nil}}$};
\draw[dotted, thick, ->](1.5,4.4)--(1.5,3.7);
\node at (1.7,4.1){$\boldsymbol{1}$};
\node at (1.5,3.5){$\boldsymbol{\mathrm{nil}}$};
\draw[dotted, thick, ->](3.5,4.4)--(3.5,3.7);
\node at (3.7,4.1){$\boldsymbol{1}$};
\node at (3.5,3.5){$\boldsymbol{\mathrm{nil}}$};
\draw[dotted, thick, ->](4.5,4.4)--(4.5,3.7);
\node at (4.7,4.1){$\boldsymbol{1}$};
\node at (4.5,3.5){$\boldsymbol{\mathrm{nil}}$};
\draw[dotted, thick,->] (5.5,4.4)--(5.5,3.7);
\node at (5.7,4.1){$\boldsymbol{1}$};
\node at (5.5,3.5){$\boldsymbol{\mathrm{nil}}$};
\draw[dotted, thick,->] (7.5,4.4)--(7.5,3.7);
\node at (7.7,4.1){$\boldsymbol{1}$};
\node at (7.5,3.5){$\boldsymbol{\mathrm{nil}}$};
\end{tikzpicture}
\end{center}
\caption{\label{fig:ex_dist} The bisimilarity metric assigns distance $\dist(s,s') = \frac{3}{4}\lambda$ to processes $s,s'\in \proc$.}
\end{figure}
Assume a $1$-bounded pseudometric $d$ with $d(\mathrm{nil},\mathrm{nil}) = d(s_1,s_3) = d(s_1,s_4) = 0$.
It is then immediate to see that $\Bisimulation(d)(s_1,s_3) = \Bisimulation(d)(s_1,s_4) = 0$ and $\Bisimulation(d)(s_1,s_2) = \Bisimulation(d)(s_1,s_5) = 1$.
Furthermore, let $\der{s',a}=\{\pi_1,\pi_2\}$ with $\pi_1 = \frac{3}{4} \delta_{s_2} + \frac{1}{4} \delta_{s_3}$ and $\pi_2 = \frac{1}{2} \delta_{s_4} + \frac{1}{2} \delta_{s_5}$.
Then we have that $\Kantorovich(d)(\delta_{s_1},\pi_1) = \frac{3}{4}$ by the matching $\w_1 \in \W(\delta_{s_1},\pi_1)$ defined by $\w(s_1,s_2) = \frac{3}{4}$ and $\w(s_1,s_3) = \frac{1}{4}$.
Analogously, we obtain $\Kantorovich(d)(\delta_{s_1},\pi_2) = \frac{1}{2}$ by the matching $\w_2 \in \W(\delta_{s_1},\pi_2)$ defined by $\w(s_1,s_4) =\frac{1}{2}$ and $\w(s_1,s_5) = \frac{1}{2}$.
Then the Hausdorff lifting allows us to capture the distance between the nondeterministic choices in the sense that, since $s$ has a unique choice, the nondeterministic evolution of $s'$ through the leftmost or the rightmost branch determines the distance between $s'$ and $s$, namely $\Bisimulation(d)(s,s') = \max \{\frac{3}{4}\lambda, \frac{1}{2}\lambda\}= \frac{3}{4} \lambda$.
Hence, the $1$-bounded pseudometric $d$ is a bisimulation metric if it satisfies $d(s_1,s_2) = d(s_1,s_5) = 1$ and $d(s,s') \ge \frac{3}{4}\lambda$.
Furthermore the bisimilarity metric, as the fixed point of functional $\Bisimulation$, assigns to processes $s,s'$ the distance $\dist(s,s') = \frac{3}{4}\lambda$.
\qed
\end{example}

The kernel of $\dist$ is the probabilistic bisimulation, namely bisimilar processes are at distance 0.
\begin{proposition}[\!\protect{\cite{BW01b}}]
\label{prop:kernel_bis_metric}
For processes $s,t \in\proc$, $\dist(s,t)=0$ if and only if $s \sim t$.
\end{proposition}

The functional $\Bisimulation$ allows us to define a notion of distance between processes that considers only the first $k$ trasnsition steps. 

\begin{definition}
[Up-to-$k$ bisimilarity metric] 
\label{def:bisim_metric_uptok}
We define the \emph{up-to-$k$ bisimilarity metric} $\distk{k}$ for $k \in \N$ by $\distk{k} = \Bisimulation^k(\botd)$.
\end{definition}

Due to the continuity of the lifting functionals $\Kantorovich$ and $\Hausdorff$, we can infer that also the functional $\Bisimulation$ is continuous, besides monotone, thus ensuring that the closure ordinal of $\Bisimulation$ is $\omega$ \cite{vB12}. 
Hence, the up-to-$k$ bisimilarity metrics converge to the bisimilarity metric when $k \to \infty$. 

\begin{proposition}
[\!\protect{\cite{vB12}}]
\label{prop:lim_bisimd_is_bisim}
Assume an image-finite PTS such that for each transition $s \trans[a] \pi$ we have that the probability distribution $\pi$ has finite support. 
Then $\dist = \lim_{k \to \infty} \distk{k}$.
\end{proposition}


\section{The modal logic $\logic$}
\label{sec:logic}

We introduce the \emph{modal logic} $\logic$ of \cite{DD11}, which extends HML \cite{HM85} with a probabilistic choice modality that allows us to express the behavior of probabilistic distributions over processes.

\begin{definition}
[Modal logic $\logic$, \cite{DD11}]
\label{def:logic}
The classes of \emph{state formulae} $\logicstate$ and \emph{distribution formulae} $\logicdist$ over $\Act$ are defined by the following BNF-like grammar: 
\[ \displaystyle
\begin{array}{rcl} 
\logicstate\colon & \varphi ::= & 
                \top  \ |\  
                 \neg \varphi   \ |\ 
                 \displaystyle \bigwedge_{j \in J}\varphi_j
                 \ | \ \diam{a}\psi \\[0.5ex]

\logicdist\colon & \psi ::= & 
\displaystyle \bigoplus_{i\in I}r_i\varphi_i  
\end{array}
\]
where: 
\begin{inparaenum}[(i)]
\item 
$\varphi$ ranges over $\logicstate$, 
\item
$\psi$ ranges over $\logicdist$,
\item 
$a\in\Act$,
\item $J \neq \emptyset$ is an at most countable set of indexes,
\item $I \neq \emptyset$ is a finite set of indexes and
\item for all $i\in I$, we have $r_i\in (0,1]$, and $\sum_{i\in I} r_i=1$.
\end{inparaenum}
\end{definition}

We shall write $\varphi_1 \wedge \varphi_2$ for $\bigwedge_{j \in J} \varphi_j$ with $J = \{1,2\}$, and $\diam{a}\varphi$ for $\diam{a} \bigoplus_{i\in I}r_i\varphi_i$ with $I = \{i\}$, $r_i=1$ and $\varphi_i = \varphi$.
Notice that instead of using $\top$ we could use $\bigwedge_{\emptyset}$.
We decided to use $\top$ to improve readability.

Formulae are interpreted over a PTS. The satisfaction relation formalizes which processes satisfy state formulae and which probability distributions satisfy distribution formulae.
Notice that, by definition, given a formula  $\bigoplus_{i \in I} r_i \varphi_i$ it holds $r_i \in (0,1]$ for all $i \in I$ and $\sum_{i \in I} r_i = 1$.
Hence, we can see the distribution formula $\bigoplus_{i \in I} r_i \varphi_i$ as a probability distribution over the state formulae $\varphi_i$. 
Thus, the formulae of the form $\diam{a}\bigoplus_{i \in I}r_i \varphi_i$ allow us to naturally capture the state-to-distribution transitions, $s \trans[a] \pi$, of PTSs.

\begin{definition}
[Satisfaction relation, \cite{DD11}]
\label{def:satisfiability}
The \emph{satisfaction relation} $\models \, \subseteq (\proc \times \logicstate) \cup (\ProbDist{\proc} \times \logicdist)$ is defined by structural induction on state formulae in $\logicstate$ by
\begin{itemize}
\item $s \models \top$ always;
\item $s \models \neg\varphi$ if{f} $s \models \varphi$ does not hold;
\item $ \displaystyle s \models \bigwedge_{j \in J} \varphi_j$ if{f} $s \models \varphi_j$ for all $j \in J$;
\item $s \models \diam{a}\psi$ if{f} $s \trans[a] \pi$ for a distribution $\pi \in \ProbDist{\proc}$ with $\pi\models \psi$,
\end{itemize}
and on distribution formulae in $\logicdist$ by
\begin{itemize}
\item $\displaystyle \pi \models \bigoplus_{i\in I}r_i\varphi_i$ if{f} $\displaystyle \pi = \sum_{i \in I}r_i \pi_i$ for some distributions $\pi_i \in \ProbDist{\proc}$ such that for all $i \in I$ we have $s \models \varphi_i$ for all states $s \in \support(\pi_i)$.
\end{itemize}
\end{definition}

\begin{example}
\label{ex:satisfiability}
Consider process $s \in \proc$ represented in Figure~\ref{fig:ex_bisim}.
It is then immediate to infer that $s \models \diam{a}\diam{b}\top$ since by executing action $a$ process $s$ evolves to the Dirac distribution on process $s_1$ which then executes action $b$.
However, we remark that we have also $s \models \diam{a}\big( \frac{3}{4} (\diam{b}\top \wedge \diam{b}\top) \oplus \frac{1}{4} \diam{b} \top \big)$.
In fact, as already noticed in Example~\ref{ex:bisim}, we can rewrite the Dirac distribution $\delta_{s_1}$ as the convex combination $\delta_{s_1} = \frac{3}{4} \delta_{s_1} + \frac{1}{4} \delta_{s_1}$ and since clearly $s_1 \models \diam{b}\top$, thus implying that also $s_1 \models \diam{b}\top \wedge \diam{b}\top$ holds, we obtain that $\delta_{s_1} \models \frac{3}{4} (\diam{b}\top \wedge \diam{b}\top) \oplus \frac{1}{4} \diam{b} \top$.
As $\delta_{s_1} \in \der{s,a}$, we can conclude that $s \models \diam{a}\big( \frac{3}{4} (\diam{b}\top \wedge \diam{b}\top) \oplus \frac{1}{4} \diam{b} \top \big)$.
\qed
\end{example}

We introduce the relation of \emph{$\logic$-equivalence} between formulae in $\logic$, which identifies formulae that are indistinguishable by their structure.
\begin{definition}
[$\logic$-equivalence]
\label{def:eqlog}
The $\logic$-\emph{equivalence} $\eqlog \subseteq (\logicstate \times \logicstate) \cup (\logicdist \times \logicdist)$ is the least equivalence relation satisfying: 
\begin{itemize}
\item $\displaystyle  \bigwedge_{j \in J} \varphi_j \eqlog \bigwedge_{j \in (J \setminus \{h\}) \cup I} \varphi_j$ whenever $\varphi_h = \displaystyle  \bigwedge_{i \in I} \varphi_i$, for $I \cap J = \emptyset$; 
\item $\displaystyle  \bigwedge_{j \in J} \varphi_j \eqlog \bigwedge_{j \in (J \setminus \{i\})} \varphi_j$ whenever $\varphi_i = \displaystyle  \bigwedge_{j \in I} \varphi_j$, for $I \subseteq J \setminus \{i\}$; 
\item $\displaystyle \bigwedge_{j \in J} \varphi_j \eqlog \bigwedge_{i \in I} \varphi_i$ whenever there is a bijection $f \colon J \to I$ with $\varphi_j \eqlog \varphi_{f(j)}$ for all $j \in J$; 
\item $\neg \varphi \eqlog \neg \varphi'$ whenever  $\varphi \eqlog \varphi'$;
\item $\diam{a}\psi_1 \eqlog \diam{a}\psi_2$ whenever $\psi_1 \eqlog \psi_2$;
\item $\bigoplus_{i \in I} r_i \varphi_i \eqlog \bigoplus_{i \in I \atop j_i \in J_i } r_{j_i} \varphi_{j_i} \text{ if{f} } \sum_{j_i \in J_i}r_{j_i} = r_i \text{ and } \varphi_{j_i} \eqlog \varphi_i \text{ for all } j_i \in J_i$.
\end{itemize}
\end{definition}

\begin{example}
\label{ex:eqlog}
Consider the state formulae $\varphi_1 = \diam{a}\diam{b}\top$ and $\varphi_2 = \varphi_1 \wedge \varphi_3$ for $\varphi_3 = \diam{a}\big( \frac{3}{4} (\diam{b}\top \wedge \diam{b}\top) \oplus \frac{1}{4} \diam{b} \top \big)$.
It is quite immediate to see that $\varphi_1 \eqlog \varphi_3$.
In fact as by Definition~\ref{def:eqlog} we have $\diam{b}\top \eqlog \diam{b}\top \wedge \diam{b}\top$ we directly obtain that the distribution formula $1\diam{b}\top$ is $\logic$-equivalent to the distribution formula $\frac{3}{4} (\diam{b}\top \wedge \diam{b}\top) \oplus \frac{1}{4} \diam{b} \top$.
Intuitively, $\varphi_1$ being $\logic$-equivalent to both the formulae defining $\varphi_2$ should guarantee the $\logic$-equivalence between $\varphi_1$ and $\varphi_2$.
To see this, we exploit Definition~\ref{def:eqlog} obtaining that $\varphi_1 \eqlog \varphi_1 \wedge \varphi_1$.
Then we gather that $\varphi_1 \wedge \varphi_1 \eqlog \varphi_2$ since we can rewrite $\varphi_1 \wedge \varphi_1$ as the state formula $\varphi_1^1 \wedge \varphi_1^2$ for $\varphi_1^1 = \varphi_1^2 = \varphi_1$ and then define the bijection $f \colon \{\varphi_1^1,\varphi_1^2\} \to \{\varphi_1, \varphi_3\}$ as $f(\varphi_1^1) = \varphi_1$ and $f(\varphi_1^2) = \varphi_3$ so that $\varphi_1^1 \eqlog f(\varphi_1^1)$ and $\varphi_1^2 \eqlog f(\varphi_1^2)$.
Hence, by the transitivity of the relation $\eqlog$, we can conclude that $\varphi_1 \eqlog \varphi_2$.
\qed
\end{example}

We can prove  that the $\logic$-equivalence respects the satisfaction relation.
\begin{proposition}
\label{prop:struc_equiv}
\begin{enumerate}
\item
If $\varphi \eqlog \varphi'$ then for all processes $s \in \proc$ we have that $s \models \varphi$ if{f} $s \models \varphi'$.
\item
If $\psi \eqlog \psi'$ then for all distributions $\pi \in \ProbDist{\proc}$ we have that $\pi \models \psi$ if{f} $\pi \models \psi'$.
\end{enumerate}
\end{proposition}

Finally, we introduce the notion of \emph{modal depth} of a formula.
Intuitively, the depth of a formula is defined as the maximum number of nested occurrences of the diamond modality in the formula. 
\begin{definition}
[Modal depth]
\label{def:depth}
The \emph{modal depth} $\depth{\varphi}$ is defined inductively for $\varphi \in \logicstate$ by
\begin{inparaenum}[(i)]
\item $\depth{\top}=0$;
\item $\depth{\neg \varphi}= \depth{\varphi}$;
\item $\depth{\bigwedge_{j \in J} \varphi_j} = \sup_{j \in J}\, \depth{\varphi_j}$;
\item $\depth{\diam{a} \psi} = 1 + \depth{\psi}$,
\end{inparaenum}
and the modal depth $\depth{\psi}$ is defined for $\psi \in \logicdist$ by
\begin{inparaenum}[(i)]
\setcounter{enumi}{4}
\item
$\depth{ \bigoplus_{i\in I}r_i\varphi_i} = \max_{i\in I}\, \depth{\varphi_i}$.
\end{inparaenum}
\end{definition}


\section{$\logic$-characterization of probabilistic bisimilarity}
\label{sec:bis_char}

In this section we introduce the notion of \emph{mimicking formula} for a process $s \in \proc$ as a formula capturing the branching and probabilistic features of $s$.
Mimicking formulae weak expressively characterize probabilistic bisimilarity: two processes are bisimilar if and only if their mimicking formulae are $\logic$-equivalent (Theorem~\ref{thm:strong_characterization}).
Moreover, we prove that the mimicking formula of a process coincides with the characteristic formula of that process w.r.t.\ probabilistic ready simulation thus allowing for an expressive characterization of the preorder: a process $t$ satisfies the mimicking formula of process $s$ if and only if $t$ ready simulates $s$ (Theorem~\ref{thm:ready_sim_char}).
Finally, we obtain the characteristic formulae for probabilistic simulation from the negation free subformulae of the mimicking formulae, thus obtaining an expressive characterization of similarity: a process $t$ satisfies the \emph{simulation characteristic formula} of process $s$ if and only if $t$ simulates $s$ (Theorem~\ref{thm:characteristic_sim}).

Mimicking formulae are defined inductively over the depth of formulae as \emph{up-to-$k$ mimicking formulae}.
Intuitively, an \emph{up-to-$k$ mimicking formula} of process $s$, denoted by $\varphi_s^k$, characterizes the branching structure of the first $k$-steps of $s$ by specifying which transitions are enabled for $s$ as well as all the actions that it cannot perform. 
Moreover, as states evolve into distributions, the up-to-$k$ mimicking formula of $s$ captures the probabilistic features of the process via the \emph{up-to-($k$-$1$) mimicking distribution formulae}. 

\begin{definition}
[Mimicking formula]
\label{def:char_formula}
For a process $s \in \proc$ and $k \in \N$, the \emph{up-to-$k$ mimicking formula of} $s$, notation $\varphi_s^k$, is defined inductively by
\begin{align*}
\varphi_s^0 & =  \top \\
\varphi_s^{k} & =  \bigwedge_{\sgotopi}\diam{a}\psi_{\pi}^{k-1} \wedge \bigwedge_{\snotdob} \neg \diam{b}\top
\end{align*}
and for a distribution $\pi \in \ProbDist{\proc}$ the \emph{up-to-$k$ mimicking distribution formula of} $\pi$, notation $\psi_{\pi}^k$, is defined by
\begin{align*}
\psi_{\pi}^k = \bigoplus_{t \in \support(\pi)} \pi(t) \varphi_{t}^k.
\end{align*}
Then, for a finite process $s \in \proc$, the \emph{mimicking formula} $\varphi_s$ is defined as $\varphi_s^k$ for $k = \depth{s} +1$.
\end{definition} 

Notice that for a finite process $s$, all formulae $\varphi_s^k$ with $k \ge \depth{s}+1$ coincide with $\varphi_s$.
Moreover, $\depth{\varphi_s} = \depth{s}+1$.  

\begin{example}
\label{ex:mimicking}
For simplicity, we assume the set $\Act = \{a,b,c\}$ as the set of actions.
Furthermore, for each $a \in \Act$, we let $\bar{a}$ denote the formula $\neg\diam{a}\top$. 
Consider the processes $s,t \in \proc$ in Figure~\ref{fig:ex_bisim} and process $s' \in \proc$ in Figure~\ref{fig:ex_dist}.
Then their mimicking formulae are defined as follows:
\begin{align*}
\varphi_s = & \diam{a}\Big(\diam{b}(\bar{a} \wedge \bar{b} \wedge \bar{c}) \wedge \bar{a} \wedge \bar{c} \Big) \wedge \bar{b} \wedge \bar{c} \\
\varphi_t = & \diam{a} \bigg( \frac{3}{4} \Big( \diam{b}(\bar{a} \wedge \bar{b} \wedge \bar{c}) \wedge \diam{b}(\bar{a} \wedge \bar{b} \wedge \bar{c}) \wedge \bar{a} \wedge \bar{c} \Big) \oplus \frac{1}{4} \Big( \diam{b}(\bar{a} \wedge \bar{b} \wedge \bar{c}) \wedge \bar{a} \wedge \bar{c} \Big)\bigg) \\
& \wedge \diam{a}\Big(\diam{b}(\bar{a} \wedge \bar{b} \wedge \bar{c}) \wedge \bar{a} \wedge \bar{c} \Big) \wedge \bar{b} \wedge \bar{c} \\
\varphi_{s'} = & \diam{a} \bigg( \frac{3}{4} \Big( \diam{b}(\bar{a} \wedge \bar{b} \wedge \bar{c}) \wedge \diam{c}(\bar{a} \wedge \bar{b} \wedge \bar{c}) \wedge \bar{a} \Big) \oplus \frac{1}{4} \Big(\diam{b}(\bar{a} \wedge \bar{b} \wedge \bar{c}) \wedge \bar{a} \wedge \bar{c} \Big) \bigg) \\
& \wedge \diam{a} \bigg( \frac{1}{2} \Big( \diam{b}(\bar{a} \wedge \bar{b} \wedge \bar{c}) \wedge \bar{a} \wedge \bar{c} \Big) \oplus \frac{1}{2} \Big( \diam{c}(\bar{a} \wedge \bar{b} \wedge \bar{c}) \wedge \bar{a} \wedge \bar{b} \Big)\bigg) \wedge \bar{b} \wedge \bar{c}.
\end{align*} 
\qed
\end{example}

As expected, each process satisfies its own mimicking formula.
\begin{theorem}
\label{thm:sat_char_form}
Given any process $s \in \proc$, $s \models \varphi_s^k$ for all $k \in \N$. Moreover, if s is finite, then $s \models \varphi_s$.
\end{theorem}

\begin{example}
\label{ex:thm1}
Consider process $s \in\proc$ represented in Figure~\ref{fig:ex_bisim}.
We have already discussed in Example~\ref{ex:satisfiability} that $s$ satisfies the formula $\diam{a}\diam{b}\top$.
Now we notice that since process $\mathrm{nil}$ cannot execute any action, clearly we have that $\mathrm{nil} \models \bar{a} \wedge \bar{b} \wedge \bar{c}$, thus giving $s_1 \models \diam{b}(\bar{a} \wedge \bar{b} \wedge \bar{c})$ and therefore $s \models \diam{a}\diam{b}(\bar{a} \wedge \bar{b} \wedge \bar{c})$.
Since moreover the only action that process $s_1$ can perform is $b$, we can infer that $s_1 \models \bar{a} \wedge \bar{c}$, thus implying $s \models \diam{a}\big( \diam{b}(\bar{a} \wedge \bar{b} \wedge \bar{c}) \wedge \bar{a} \wedge \bar{c} \big)$.
Finally, since $s$ can execute neither $b$ nor $c$ we have $s \models \bar{b} \wedge \bar{c}$ thus giving $s \models \varphi_s$, where $\varphi_s$ is the mimicking formula of $s$ represented in Example~\ref{ex:mimicking}.
\qed
\end{example}

Mimicking formulae allow us to characterize probabilistic bisimilarity. 
\begin{theorem}
\label{thm:strong_characterization}
Given any processes $s, t \in \proc$ and $k \in \N$, $\varphi_s^k \eqlog \varphi_t^k$ if and only if $s \sim_k t$. Moreover, if $s$ and $t$ are finite, $\varphi_s \eqlog \varphi_t$  if and only if $s \sim t$.
\end{theorem}

\begin{example}
\label{ex:thm2}
As previously discussed in Example~\ref{ex:bisim}, processes $s,t \in \proc$ represented in Figure~\ref{fig:ex_bisim} are bisimilar.
We can show that their mimicking formulae, resp.\ $\varphi_s$ and $\varphi_t$ represented in Example~\ref{ex:mimicking} are $\logic$-equivalent.
From Definition~\ref{def:eqlog} to show that $\varphi_s \eqlog \varphi_t$ it is enough to show that 
\begin{align*}
\varphi_s \eqlog \varphi_1 ={} & \diam{a}\Big(\diam{b}(\bar{a} \wedge \bar{b} \wedge \bar{c}) \wedge \bar{a} \wedge \bar{c} \Big) \wedge \bar{b} \wedge \bar{c}\\
\varphi_s \eqlog \varphi_2 ={} & \diam{a} \bigg( \frac{3}{4} \Big( \diam{b}(\bar{a} \wedge \bar{b} \wedge \bar{c}) \wedge \diam{b}(\bar{a} \wedge \bar{b} \wedge \bar{c}) \wedge \bar{a} \wedge \bar{c} \Big) \oplus \frac{1}{4} \Big( \diam{b}(\bar{a} \wedge \bar{b} \wedge \bar{c}) \wedge \bar{a} \wedge \bar{c} \Big)\bigg).
\end{align*}
We stress that, since $\varphi_t \eqlog \varphi_1 \wedge \varphi_2$, from $\varphi_s \eqlog \varphi_1$ we can infer that process $s$ can match the rightmost branch of process $t$ and that from $\varphi_s \eqlog \varphi_2$ we infer that $s$ matches the leftmost branch of $t$. 
Symmetrically, $\varphi_s \eqlog \varphi_1$ and $\varphi_s \eqlog \varphi_2$ imply that $t$ can match the $a$-labeled transition of $s$ through both branches.
The $\logic$-equivalence between $\varphi_s$ and $\varphi_1$ is trivial since we have $\varphi_s = \varphi_1$ (cf.\ Example~\ref{ex:mimicking}).
Next we notice that if we consider the negation free versions of formulae $\varphi_s$ and $\varphi_2$, namely if we do not consider the occurrences of formulae of the form $\bar{a}$ for $a \in \Act$, then the $\logic$-equivalence between the two state formulae is given in Example~\ref{ex:eqlog}.
Since the negated formulae are the same and their occurrences in the structure of $\varphi_s$ and $\varphi_2$ are the same, we can directly conclude that $\varphi_s \eqlog \varphi_2$.
\qed
\end{example}

Mimicking formulae capture all possible resolutions of nondeterminism of processes as well as their inability to perform a specific action.
Consequently, they give us enough power to expressively characterize ready simulation: the mimicking formula of a process $s$ is the characteristic formula of $s$ w.r.t.\ ready simulation. 

\begin{theorem}
\label{thm:ready_sim_char}
Given any processes $s,t \in \proc$ and $k \in \N$, $t \models \varphi_s^k$  if and only if  $s \,\sqsubseteq_k^r\, t$.
Moreover, if $s$ is finite, then  $t \models \varphi_s$ if and only if  $s \,\sqsubseteq^r\, t$.
\end{theorem}

We aim to obtain a similar result for simulation.
Firstly, we notice that whenever a process $t$ satisfies the mimicking formula $\varphi_s$ of process $s$ we are guaranteed that all transitions performed by $s$ are mimicked by transitions by $t$.
Thus, the following soundness results with respect to simulation is natural.

\begin{theorem}
\label{thm:sim_char}
Given any processes $s, t \in \proc$ and $k \in \N$,  if $t \models \varphi_s^k$ then $s \,\sqsubseteq_k\, t$.
Moreover, if $s$ is finite it holds that whenever $t \models \varphi_s$ then $s \,\sqsubseteq\, t$.
\end{theorem}

The distinguishing power of mimicking formulae is too strong to obtain completeness: a process $s$ with $\init{s} = \emptyset$ is simulated by any process $t$, but the mimicking formula of $s$, $\varphi_s = \bigwedge_{a \in \Act} \neg \diam{a}\top$, is satisfied only by those $t$ with $\init{t} = \emptyset$.
However, if we consider the negation free subformula of a mimicking formula then we obtain the characteristic formula for simulation of a process (Theorem~\ref{thm:characteristic_sim}).

\begin{definition}
[Simulation characteristic formula]
\label{def:characteristic}
For a process $s \in \proc$ and $k \in \N$, the \emph{up-to-$k$ simulation characteristic formula of} $s$, notation $\vartheta^k_s$, is defined inductively by:
\begin{align*}
\vartheta_s^0 & =  \top \\
\vartheta_s^{k} & =  \bigwedge_{\sgotopi}\diam{a} \upsilon_{\pi}^{k-1}
\end{align*}
and for a distribution $\pi \in \ProbDist{\proc}$ the \emph{up-to-$k$ simulation characteristic distribution formula of} $\pi$, notation $\upsilon_{\pi}^k$, is defined by
\begin{align*}
\upsilon_{\pi}^k = \bigoplus_{t \in \support(\pi)} \pi(t) \vartheta_{t}^k.
\end{align*}
Then, for a finite process $s \in \proc$, the \emph{simulation characteristic formula} $\vartheta_s$ is $\vartheta_s^k$ for $k = \depth{s}$.
\end{definition}

Notice that for a finite process $s$, all formulae $\vartheta_s^k$ with $k \ge \depth{s}$ coincide with $\vartheta_s$.
We remark that for each finite process $s$ we have $\depth{\varphi_s} = \depth{\vartheta_s} + 1$ since the mimicking formula of a process with no outgoing transitions is $\bigwedge_{a \in \Act} \neg\diam{a} \top$, whose depth is $1$, whereas the simulation characteristic formula of the same process is $\top$, whose depth is $0$.

\begin{theorem}
\label{thm:characteristic_sim}
Given processes $s,t \in \proc$ and $k \in \N$, $t \models \vartheta_s^k$ if and only if $s \,\sqsubseteq_k\, t$.
Moreover, if $s$ is finite, then $t \models \vartheta_s$ if and only if $s \,\sqsubseteq\, t$.
\end{theorem}


\section{$\logic$-characterization of bisimilarity metric}
\label{sec:metric_char}

In this section we present the main contribution of this paper: the logical characterization of bisimilarity metric. 
We provide a suitable distance between formulae in $\logic$ and we characterize bisimilarity metric as the distance between the mimicking formulae of processes (Theorem~\ref{thm:ell=bisim}).

As previously outlined, distribution formulae can be considered as probability distributions over state formulae.
It is then natural to adapt the notion of matching for probability distributions to a \emph{logical matching for distribution formulae}.
\begin{definition}
[Logical matching]
\label{def:matching_oplus}
A \emph{logical matching} $\w$ for distribution formulae $\psi_1 = \bigoplus_{i \in I} r_i \varphi_i$ and $\psi_2 = \bigoplus_{j \in J} r_j \varphi_j$ is a probability distribution $\w$ over the product space $\logicstate \times \logicstate$ with $\psi_1$ and $\psi_2$ as left and right marginal, resp., that is $\sum_{\varphi \in \logicstate} \w(\varphi_i,\varphi) = r_i$ and $\sum_{\varphi \in \logicstate} \w(\varphi,\varphi_j) = r_j$, for all $i \in I$ and $j \in J$. 
We denote by $\W(\psi_1,\psi_2)$ the set of all logical matchings for $\psi_1$ and $\psi_2$.
\end{definition}

Next, we introduce the distance between formulae which is defined inductively over the depth of formulae and their structure. \begin{definition}
[Up-to-$k$ distance between formulae]
\label{def:distfs}
Let $\lambda \in (0,1]$ be a discount factor.
For $k \in \N$, the 
\emph{up-to-$k$ distance between state formulae} is the mapping $\distfsk{k} \colon \logicstate \times \logicstate \to [0,1]$ defined by
\begin{itemize}
\item
$\distfsk{0}(\varphi_1,\varphi_2)  = 0$ for all  $\varphi_1,\varphi_2 \in \logicstate$
\item
$\distfsk{k}(\varphi_1, \varphi_2)  =  \begin{cases}
                                                            0 & 
                                                                   \text{if } \varphi_1=\top \text{ and } \varphi_2=\top\\[.3ex]
                                                            \distfsk{k}(\varphi'_1, \varphi'_2) & 
                                                                   \text{if } \varphi_1 = \neg\varphi_1' \text{ and }  \varphi_2 = \neg\varphi'_2\\[.3ex]
                                                            \lambda \cdot \distfdk{k-1}(\psi_1,\psi_2) & 
                                                                   \text{if } \varphi_1 = \diam{a}\psi_1 \text{ and } \varphi_2 = \diam{a}\psi_2\\[.4ex]
                                                            \max \left\{ \begin{array}{c}
                                                                                \displaystyle{\sup_{j \in J}\, \inf_{i \in I}\, \distfsk{k}(\varphi_j,\varphi_i),} \\[.3ex]
                                                                               \displaystyle{\sup_{i \in I}\, \inf_{j \in J}\, \distfsk{k}(\varphi_j,\varphi_i)}
                                                                               \end{array} \right\} & 
                                                                   \text{if } \left(
                                                                                  \begin{array}{c}  \displaystyle{\varphi_1 = \bigwedge_{j \in J} \varphi_j} \text{ and }  
                                                                                    \displaystyle{\varphi_2 = \bigwedge_{i \in I} \varphi_i}, \\
                                                                                     \text{ with } \sup \emptyset = 0 \text{ and }\inf \emptyset = 1
                                                                                  \end{array} \right) \\[.3ex]
                                                            1 & 
                                                                   \text{otherwise }\\
                                                            \end{cases}$
\end{itemize}
and the \emph{up-to-$k$ distance between distribution formulae} $\distfdk{k} \colon \logicdist \times \logicdist \to [0,1]$ is defined by
\begin{itemize}
\item $\displaystyle \distfdk{k}(\psi_1,\psi_2) = \min_{\w \in \W(\psi_1,\psi_2)} \sum_{\varphi',\varphi'' \in \logicstate} \w(\varphi',\varphi'')  \cdot \distfsk{k}(\varphi',\varphi'')$.
\end{itemize}
\end{definition}

The discount factor $\lambda \in (0,1]$ has a similar purpose to that used in the definition of the bisimulation metric functional $\Bisimulation$ (Definition~\ref{def:metric_bisim_functional}), by reasoning on an \emph{in depth} manner.
It allows us to specify how much the distance between state formulae at the same depth is taken into account.
For this reason, the discount factor $\lambda$ is introduced in the evaluation of the distance between equally labeled diamond modalities.

\begin{example}
\label{ex:dist_formulae}
We aim to evaluate, for $k = 1,2,3$, the up-to-$k$ distance between the state formulae 
$\varphi_1 = \diam{a}\psi_1 \wedge \bar{b} \wedge \bar{c}$ and 
$\varphi_2 = \diam{a}\psi_2 \wedge \bar{b} \wedge \bar{c}$ where 
$\psi_1 = 1 \phi_1$ with $\phi_1 = \diam{b}(\bar{a} \wedge \bar{b} \wedge \bar{c}) \wedge \bar{a} \wedge \bar{c}$ and $\psi_2 = \frac{3}{4} \phi_2 \oplus \frac{1}{4} \phi_3$ with $\phi_2 = \diam{b}(\bar{a} \wedge \bar{b} \wedge \bar{c}) \wedge \diam{c}(\bar{a} \wedge \bar{b} \wedge \bar{c}) \wedge \bar{a}$ and $\phi_3 = \diam{b}(\bar{a} \wedge \bar{b} \wedge \bar{c}) \wedge \bar{a} \wedge \bar{c}$.
By Definition~\ref{def:distfs} we have
\begin{align*}
\distfsk{k}(\varphi_1,\varphi_2) ={} & 
\max 
\left\{ 
	\begin{array}{l} 
	\max 
	\left\{
		\begin{array}{l}
		\min \{ \distfsk{k}(\diam{a}\psi_1,\diam{a}\psi_2), \distfsk{k}(\diam{a}\psi_1, \bar{b}), \distfsk{k}(\diam{a}\psi_1,\bar{c})\} \\
		\min \{ \distfsk{k}(\bar{b},\diam{a}\psi_2), \distfsk{k}(\bar{b}, \bar{b}), \distfsk{k}(\bar{b},\bar{c})\} \\				
		\min \{ \distfsk{k}(\bar{c},\diam{a}\psi_2), \distfsk{k}(\bar{c}, \bar{b}), \distfsk{k}(\bar{c},\bar{c})\}				
		\end{array} 
	\right\} \\
	\max 
	\left\{
		\begin{array}{l}
		\min \{ \distfsk{k}(\diam{a}\psi_1,\diam{a}\psi_2), \distfsk{k}(\bar{b},\diam{a}\psi_2), \distfsk{k}(\bar{c},\diam{a}\psi_2)\} \\
		\min \{ \distfsk{k}(\diam{a}\psi_1, \bar{b}), \distfsk{k}(\bar{b}, \bar{b}), \distfsk{k}(\bar{c},\bar{b})\} \\				
		\min \{ \distfsk{k}(\diam{a}\psi_1,\bar{c}), \distfsk{k}(\bar{b}, \bar{c}), \distfsk{k}(\bar{c},\bar{c})\}				
		\end{array} 
	\right\} \\
	\end{array}
\right\} \\
={} &
\max 
\left\{ 
	\begin{array}{l} 
	\max 
	\left\{
		\begin{array}{l}
		\min \{ \distfsk{k}(\diam{a}\psi_1,\diam{a}\psi_2), 1, 1\} \\
		\min \{ 1, 0, 1\} \\				
		\min \{ 1, 1, 0\}				
		\end{array} 
	\right\} \\
	\max 
	\left\{
		\begin{array}{l}
		\min \{ \distfsk{k}(\diam{a}\psi_1,\diam{a}\psi_2), 1, 1\} \\
		\min \{ 1, 0, 1\} \\				
		\min \{ 1, 1, 0\}				
		\end{array} 
	\right\} \\
	\end{array}
\right\} \\
={} & \distfsk{k}(\diam{a}\psi_1,\diam{a}\psi_2) \\
={} & \lambda \cdot \distfdk{k-1}(\psi_1,\psi_2) \\
={} & \lambda \cdot \min_{\w \in \W(\psi_1,\psi_2)} \sum_{i \in \{2,3\}} \w(\phi_1,\phi_i)  \cdot \distfsk{k}(\phi_1,\phi_i)
\end{align*}
where the second equality follows from Definition~\ref{def:distfs} which gives that, for all $a \in \Act$, $\distfsk{k}(\bar{a},\varphi) = 0$ if $\varphi= \bar{a}$ and $\distfsk{k}(\bar{a},\varphi) = 1$ in all other cases.
It is quite immediate to verify that for $k = 1$ we have $\distfsk{1}(\varphi_1,\varphi_2) = 0$. 
In fact by definition of up-to-$0$ distance between formulae we have that $\distfsk{0}(\phi_1,\phi_i) = 0$ for all $i \in \{2,3\}$, from which we gather $\lambda \cdot \distfdk{0}(\psi_1,\psi_2) = 0$ and thus $\distfsk{1}(\varphi_1,\varphi_2) = 0$.

Consider now the case of $k = 2$.
Since $\phi_1$ and $\phi_3$ represent the same state formula we can directly infer that $\distfsk{1}(\phi_1,\phi_3) = 0$.
Moreover we have $\distfsk{1}(\phi_1,\phi_2) = 1$ due to fact the subformula $\diam{c}(\bar{a} \wedge \bar{b} \wedge \bar{c})$ in $\phi_2$ cannot be matched by any subformula of $\phi_1$.
Then we have $\distfdk{1}(\psi_1,\psi_2) = \frac{3}{4}$ by the logical matching $\w \in \W(\psi_1,\psi_2)$ defined by $\w(\phi_1, \phi_2) = \frac{3}{4}$ and $\w(\phi_1,\phi_3) = \frac{1}{4}$, thus implying $\distfsk{2}(\varphi_1,\varphi_2) = \frac{3}{4} \lambda$.

Since the formulae occurring after the $\diam{b}$ in both $\psi_1$ and $\psi_2$ are identical, we can directly conclude that $\distfsk{3}(\varphi_1,\varphi_2) = \distfsk{2}(\varphi_1,\varphi_2)$.
More precisely, we have that $\distfsk{k}(\varphi_1,\varphi_2) = \frac{3}{4} \lambda$ for all $k \ge 2$.
\qed
\end{example}

We have already noticed that distribution formulae are probability distributions over $\logicstate$.
Thus, the up-to-$k$ distance between distribution formulae plays the role of a transportation lifting of the up-to-$k$ distance between state formulae.
In particular, we have that $\distfdk{k}$ mirrors the Kantorovich lifting
\[
\distfdk{k}(\psi_1,\psi_2) = 
\min_{\w \in \W(\psi_1,\psi_2)} \sum_{\varphi',\varphi'' \in \logicstate} \w(\varphi',\varphi'')  \cdot \distfsk{k}(\varphi',\varphi'') =
\Kantorovich(\distfsk{k})(\psi_1,\psi_2).
\]
Moreover, we notice that
\[
\distfsk{k}(\bigwedge_{j \in J} \varphi_j, \bigwedge_{i \in I} \varphi_i) = 
\max \Big\{ 
                    \displaystyle{\sup_{j \in J}\, \inf_{i \in I}\, \distfsk{k}(\varphi_j,\varphi_i),\;}
                    \displaystyle{\sup_{i \in I}\, \inf_{j \in J}\, \distfsk{k}(\varphi_j,\varphi_i)}
        \Big\} 
= \Hausdorff(\distfsk{k}) \Big(\{\varphi_j \mid j \in J\}, \{\varphi_i \mid i \in I\} \Big)
\]
expressing the maximum between the greatest of all distances from a $J$-indexed formula to the closest $I$-indexed formula and viceversa.
We recall that the Hausdorff lifting is used in the definition of bisimulation metrics in order to capture nondeterministic choices (Definition~\ref{def:Hausdorff}).
Here, we use it to quantify the distance between conjunctions of formulae, which is natural since in mimicking formulae the conjunction of formulae is used to capture nondeterminism.
The close relation between our distance on $\logic$ and the Hausdorff and Kantorovich pseudometrics will be crucial in the characterization of bisimilarity metric (Theorem~\ref{thm:ell=bisim}).

The mapping $\distfsk{k}$ is actually a $1$-bounded pseudometric which is preserved modulo $\logic$-equivalence.
\begin{proposition}
\label{prop:distfsk_is_metric}
For all $k  \in \N$, $\distfsk{k}$ and $\distfdk{k}$ are 1-bounded psedudometrics.
\end{proposition}

\begin{proposition}
\label{prop:distfsk_equiv}
For all $k \in \N$ it holds that
\begin{enumerate}
\item for all $\varphi \in \logicstate$, $\distfsk{k}(\varphi, \varphi') = \distfsk{k}(\varphi, \varphi'')$ for all $\varphi',\varphi'' \in \logicstate$ with $\varphi' \eqlog \varphi''$ and
\item for all $\psi \in \logicdist$, $\distfdk{k}(\psi,\psi') = \distfdk{k}(\psi,\psi'')$ for all $\psi',\psi'' \in \logicdist$ with $\psi' \eqlog \psi''$.
\end{enumerate}
\end{proposition}

We define the \emph{distance between formulae}, denoted by $\distfs$, as the limit of their up-to-$k$ distances.
The existence of such a limit is guaranteed by the two following results.

\begin{lemma}
\label{lem:distfs_monotony}
For each $k \in \N$ it holds that
\begin{enumerate}
\item for all $\varphi,\varphi' \in \logicstate$, $\distfsk{k+1}(\varphi, \varphi') \ge \distfsk{k}(\varphi, \varphi')$ and
\item for all $\psi,\psi' \in \logicdist$, $\distfdk{k+1}(\psi, \psi') \ge \distfdk{k}(\psi, \psi')$.
\end{enumerate}
\end{lemma}

\begin{proposition}
\label{prop:lim_distfs}
The mapping $\distfs \colon \logicstate \times \logicstate$ defined by 
\[ 
\distfs(\varphi, \varphi') = \lim_{k \to \infty} \distfsk{k}(\varphi, \varphi')
\] 
for all $\varphi, \varphi' \in \logicstate$, is well-defined. 
Analogously, the mapping $\distfd \colon \logicdist \times \logicdist$ defined by 
\[ 
\distfd(\psi, \psi') = \lim_{k \to \infty} \distfdk{k}(\psi, \psi')
\] 
for all $\psi, \psi' \in \logicdist$, is well-defined. 
\end{proposition}

The distance between formulae is indeed a $1$-bounded pseudometric whose kernel is $\logic$-equivalence.

\begin{proposition}
\label{prop:distfs_is_metric}
The mappings $\distfs$ and $\distfd$ are 1-bounded pseudometrics.
\end{proposition}

\begin{proposition}
\label{prop:distfs_kernel}
\begin{enumerate}
\item Let $\varphi, \varphi' \in \logicstate$.
Then $\distfs(\varphi, \varphi')=0$ if and only if $\varphi \eqlog \varphi'$.
\item Let $\psi, \psi' \in \logicdist$.
Then $\distfd(\psi, \psi')=0$ if and only if $\psi \eqlog \psi'$.
\end{enumerate}
\end{proposition}

We are ready to lift the metric on $\logic$ to a metric on $\proc$.
To this aim, we exploit the close relation between processes and their own mimicking formulae.
All distances between probabilistic processes proposed in the literature take into account the disparities in their branching structures as well as the differences between the probabilistic choices, in order to conciliate behavioral equivalence with quantitative properties.
By construction, each mimicking formula is univocally determined by the process and in turn the branching and probabilistic structure of the process are univocally captured by that formula.
Hence, we define the \emph{logical distance} on processes as the distance between their mimicking formulae.

\begin{definition}
[Logical distance]
\label{def:logic_distance}
For any $k \in \N$, the \emph{up-to-$k$ logical distance} $\ell_{\lambda}^k \colon \proc \times \proc \to [0,1]$ over processes is defined by
\[
\ell_{\lambda}^k(s,t)  = \distfsk{k}(\varphi_s^k,\varphi_t^k).
\]
for all $s,t \in \proc$.
Moreover, if $s$ and $t$ are finite then we define the \emph{logical distance} $\ell_{\lambda} \colon \proc \times \proc \to [0,1]$ as
\[
\ell_{\lambda}(s,t)  = \distfs(\varphi_s,\varphi_t).
\]
\end{definition}
Notice that if $s$ and $t$ are finite then $\ell_{\lambda}(s,t) = \lim_{k \to \infty} \ell_{\lambda}^k(s,t)$. 

\begin{proposition}
\label{prop:ell_is_metric}
\begin{enumerate}
\item For any $k \in \N$ the mapping $\ell_{\lambda}^k$ is a 1-bounded pseudometric.
\item The mapping $\ell_{\lambda}$ is a 1-bounded pseudometric.
\end{enumerate}
\end{proposition}

The next Theorem, which is our main result, states the equivalence between the up-to-$k$ logical distance and the up-to-$k$ bisimilarity metric for each $k \in \N$.

\begin{theorem}
\label{thm:ell=bisim}
Let $\lambda \in (0,1]$ be a discount factor.
Given any processes $s,t \in \proc$ and $k \in \N$, $\ell_{\lambda}^k(s,t) = \distk{k}(s,t)$.
Moreover, if $s$ and $t$ are finite, $\ell_{\lambda}(s,t) = \dist(s,t)$.
\end{theorem}

\begin{example}
\label{ex:thm6}
Consider the processes $s,s' \in \proc$ represented in Figure~\ref{fig:ex_dist}.
As argued in Example~\ref{ex:dist} we have $\dist(s,s') = \lambda \cdot \frac{3}{4}$.
Let us evaluate the logical distance between them, namely the distance between their mimicking formulae, $\varphi_s$ \and $\varphi_{s'}$ resp.\ represented in Example~\ref{ex:mimicking}.
To simplify the presentation, we rewrite $\varphi_s$ as the state formula $\diam{a} \psi_1 \wedge \bar{b} \wedge \bar{c}$ and $\varphi_{s'}$ as the state formula $\diam{a}\psi_2 \wedge \diam{a}\psi_3 \wedge \bar{b} \wedge \bar{c}$, with $\psi_1 =1 \phi_1$ for $\phi_1 = \diam{b}(\bar{a} \wedge \bar{b} \wedge \bar{c}) \wedge \bar{a} \wedge \bar{c}$, $\psi_2 = \frac{3}{4} \phi_2 \oplus \frac{1}{4} \phi_3$ for $\phi_2 = \diam{b}(\bar{a} \wedge \bar{b} \wedge \bar{c}) \wedge \diam{c}(\bar{a} \wedge \bar{b} \wedge \bar{c}) \wedge \bar{a}$ and $\phi_3 = \diam{b}(\bar{a} \wedge \bar{b} \wedge \bar{c}) \wedge \bar{a} \wedge \bar{c}$ and $\psi_3 =  \frac{1}{2} \phi_4 \oplus \frac{1}{2}\phi_5$ for $\phi_4 = \diam{b}(\bar{a} \wedge \bar{b} \wedge \bar{c}) \wedge \bar{a} \wedge \bar{c}$ and $\phi_5 = \diam{c}(\bar{a} \wedge \bar{b} \wedge \bar{c}) \wedge \bar{a} \wedge \bar{b}$.
Following the same reasoning of Example~\ref{ex:dist_formulae} we infer that $\distfs(\varphi_s,\varphi_{s'}) = \max \left\{ \distfs(\diam{a}\psi_1, \diam{a}\psi_2), \distfs(\diam{a}\psi_1, \diam{a}\psi_3) \right\}$.
Furthermore, we notice that if we consider the two state formulae $\varphi_1$ and $\varphi_2$ of previous Example~\ref{ex:dist_formulae} we gather that $\varphi_1 = \varphi_s$ and $\varphi_2 = \diam{a}\psi_2 \wedge \bar{b} \wedge \bar{c}$. 
Therefore we can directly conclude that $\distfs(\diam{a}\psi_1,\diam{a}\psi_2) = \frac{3}{4} \lambda$.
Thus, let us consider $\distfs(\diam{a}\psi_1,\diam{a}\psi_3)$.
It is clear that $\distfs(\phi_1,\phi_4) = 0$, whereas we have $\distfs(\phi_1,\phi_5) = 1$.
Hence, we obtain $\distfd(\psi_1,\psi_2) = \frac{1}{2} \lambda$ by the logical matching $\w \in \W(\psi_1,\psi_3)$ defined by $\w(\phi_1,\phi_4) = \frac{1}{2}$ and $\w(\phi_1,\phi_5) = \frac{1}{2}$, from which we can conclude that $\distfs(\varphi_s,\varphi_{s'}) = \max\{ \frac{3}{4} \lambda, \frac{1}{2} \lambda\} = \frac{3}{4} \lambda$.
We stress that using the Hausdorff lifting to evaluate the distance between conjunctions of formulae allows us to capture the nondeterministic choices in the bisimulation game (cf.\ Example~\ref{ex:dist}).
\qed
\end{example}

As an immediate consequence of Theorem~\ref{thm:ell=bisim} we obtain that bisimilarity is the kernel of the logical distance.

\begin{corollary}
\label{cor:ell=bisim}
Given any finite processes $s,t\in\proc$, $s \sim t$ if and only if $\ell_{\lambda}(s,t) = 0$.
\end{corollary}

Moreover our characterization of bisimilarity metric agrees with our characterization of bisimilarity.

\begin{corollary}
\label{cor:cor_ell=bisim}
Given any finite processes $s,t \in \proc$, $\varphi_s \eqlog \varphi_t$ if and only if $\ell_{\lambda}(s,t)=0$.
\end{corollary}

We have previously outlined that the metric $\distfs$ is the exact transposition of the Hausdorff and Kantorovich lifting functionals over the elements of $\logicstate$ and $\logicdist$, respectively.
This is one of the key features that allowed us to obtain the characterization of bisimulation metric.
So, as a final remark, we answer to the natural question that may arise: what happens if we change one or both lifting functionals in the definition of bisimilarity metric (Definition~\ref{def:metric_bisim_functional})?
In this case, the metric of Definition~\ref{def:distfs} would not be useful for the characterization result.
However the ideas exceeding the technical definition would be still valid: to obtain the logical characterization, we have to define the logical distance between processes as a suitable distance between the mimicking formulae.
Hence, if in the definition of bisimulation metric the Kantorovich lifting functional $\Kantorovich$ is substituted by another lifting functional $P$, then we should modify the distance between distribution formulae as $\distfdk{k}(\psi_1,\psi_2)=P(\distfsk{k})(\psi_1,\psi_2)$.
If conversely the Hausdorff lifting functional $\Hausdorff$ is changed with another lifting functional, then we would have to modify the definition of $\distfsk{k}$ on the boolean operator $\bigwedge$ accordingly.


\section{Conclusions, related and future work}
\label{sec:conclusion}

We have proposed a modal characterization of bisimilarity metric on finite processes based on the novel notions of mimicking formula and distance between formulae defined on the probabilistic version of HML of \cite{DD11}, $\logic$.
Mimicking formulae capture all possible resolutions of nondeterminism for the related processes by also exactly specifying the reached probability distributions.
These properties allowed for a weak expressive characterization of probabilistic bisimilarity: two processes are bisimilar if and only if their mimicking formulae are $\logic$-equivalent.
Moreover, we proved that the mimicking formula of a process $s$ coincides with the characteristic formula of $s$ w.r.t.\ probabilistic ready simulation, thus obtaining an expressive characterization of this preorder.
Finally, we showed how to derive the characteristic formula of a process w.r.t.\ probabilistic simulation from its mimicking formula.

Expressive characterizations for probabilistic equivalences and preorders are given in \cite{DvGHM08,DvG10}.
\cite{DvGHM08} deals with forward simulation \cite{S95} and probabilistic failure simulation.
The characteristic formulae for these preorders are defined on a negation free logic which is HML enriched with two operators: $\bigoplus$ to deal with probabilistic choices and $\mathbf{ref}(A)$, for $A \subseteq \Act$, expressing that actions in $A$ are not executable and thus capturing the failure semantics.
The logic used in \cite{DvGHM08} is richer than $\logic$ since it 
allows for arbitrary formulae to occur after the diamond modality. 
\cite{DvG10} extends \cite{DvGHM08} to deal with both infinite depth processes and strong and weak probabilistic (bi)simulations. 
To capture the possible infinite execution of processes, they used the probabilistic modal $\mu$-calculus which is indeed a logic much richer  than $\logic$.
However, we notice that even if we restrict the results in \cite{DvG10} to finite processes, the definition of characteristic formulae for (weak) probabilistic bisimulations would still require a  logic richer than $\logic$.
In particular, they would still need to allow arbitrary formulae occurring after the diamond modality.

Characterizations of bisimilarity metric based on $[0,1]$-valued logics are given in \cite{DGJP04,BW01b,BW05,AFS04,AMRS08,BBLM14, DDG15}.
In \cite{DGJP04} a metric between labeled Markov processes is defined by giving a real-valued semantics to a probabilistic modal logic. 
Informally, formulae are translated into functional expressions and the satisfaction relation is interpreted as integration.
Then, the distance between two processes is defined as the maximal disparity between their images through all functionals, obtaining that probabilistic bisimilar processes are the ones at distance $0$.
Later, in \cite{BW01b,BW05} it was proved that the metric in \cite{DGJP04} coincides with the bisimilarity metric based on the Kantorovich lifting.
Functional expressions are also used in \cite{AFS04} to obtain the characterization of two classes of behavioral metrics: \emph{linear distances}, capturing trace inclusion and equivalence, are characterized through the \emph{Quantitative Linear-Time Temporal Logic}, whereas the \emph{branching distances}, capturing simulation and bisimulation, are characterized with the \emph{Quantitative $\mu$-calculus} of \cite{AHM03}.
The same logic is used in \cite{AMRS08}, for stochastic game structures, to characterize \emph{a priori} metric, defined as the distance between the expected payof{f}s of the players.
Linear-time properties are also studied in \cite{BBLM14}, in order to capture approximate reasoning on Stochastic Markov Models (SMMs).
SMMs are a generalization of CTMCs in the sense that exit-time probabilities follow generic distributions on the positive real line.
For the specification of SMMs properties, they propose the \emph{Metric Temporal Logic} (MTL), built on implication and the temporal operators \emph{next} and \emph{until}.
Several equivalent distances on SMMs are proposed, one of which is the MTL-\emph{variation pseudometric} defined as the total variation distance on the probability measure on timed paths and over approximated by the convex combination of the total variation distance on the exit-time probabilities and the Kantorovich distance on transition probability functions.
Finally, in \cite{DDG15} a real-valued logic is proposed for the characterization of a \emph{state-based bisimulation metric} which coincides with the one of \cite{DCPP06} and of a \emph{distribution-based bisimulation metric} which is directly defined over distributions without using any lifting functional \cite{DFdL15,FZ14,HKK14}.

The originality of our notion of distance over  $\logic$ relies on the fact that it is not defined in terms of any ground distance between processes.
As a matter of fact, our distance between formulae is independent from the metric properties of the process space.
An example of a logical characterization obtained with a distance between formulae defined in terms of the distance on processes is in \cite{LMP12}, joint to the study of relating the behavior of approximations to the limit behavior of the system itself (approximate reasoning principles) for both discrete-time (DMPs) and continuous-time Markov processes (CMPs) with continuous state space.
To this aim, the property of \emph{dynamical continuity} for a pseudometric is introduced: a metric is dynamically continuous if it allows one to identify convergent sequences of processes or formulae.
Then, they define a metric space for the \emph{Discrete Markovian Logic} and the \emph{Continuous Markovian Logic} \cite{MCL12} by considering as distance between formulae the Hausdorff distance on the sets of processes satisfying them.
In this way, they are able to topologically characterize the logical properties induced by a dynamically continuous metric for both DMPs and CMPs.

We have already argued that we defined the distance between formulae with the exact purpose of simulating the Hausdorff and Kantorovich lifting on which the bisimilarity metric is defined. 
Despite this kind of reasoning may seem too restrictive at first glance, we believe that having a distance between formulae instead of a real-valued semantics for the logic turns out to be an advantage in case one wishes to modify the lifting functionals in the definition of bisimilarity metric (cf. last part of Section~\ref{sec:metric_char}).
Hence, we aim to extend our results to other lifting functionals, like the generalized Kantorovich lifting $\Kantorovich_V$~\cite{CGPX14}. 

Then, we aim to extend our results to infinite processes with bounded nondeterminism, that is we will allow infinite execution sequences but we will still require the image-finiteness hypothesis, since without it defining probabilistic bisimilarity as the limit of its approximations would not be possible \cite{HPSWZ11}, and the finiteness of the supports of the probability distributions, in order to guarantee the continuity of the bisimulation metric functional (cf. last part of Section~\ref{sec:background}).
To capture infinite execution we will follow the approach of \cite{L90} (later generalized in \cite{AILS12,SZ12}), also known as \emph{equational $\mu$-calculus}.
In detail, we will extend the logic $\logic$ to a \emph{modal $\proc$-indexed logic} by adding the $\proc$-indexed family of variables $\{X_{s} \mid s \in \proc\}$.
Intuitively, these variables will allow for a recursive specification of formulae and, thus, of processes.
Then, an appropriate interpretation to each variable is provided as the solution of a system of equations defined using (endo)declarations, namely functions $\E$ mapping each variable into an arbitrary formula of the logic. 
As solution of the system we will consider the variable interpretation corresponding to the greatest fixed point of the system.
Finally, we will assign to each process $s$ the related mimicking formula $\varphi_s$ defined as $\E(X_s)$.
Once we have obtained the mimicking formulae, we will be able to extend the results of this paper to infinite processes.

In the recent paper \cite{CGT16b} a SOS-based method for decomposing formulae in $\logic$ and thus deriving modal properties of nondeterministic probabilistic systems has been proposed.
In \cite{BFvG04,FvGW06,FvGW12,FvG16} the decomposition of modal formulae is used to systematically derive expressive congruence formats for several behavioral equivalences and preorders from their modal characterizations.
In the probabilistic setting, this kind of result has been given in \cite{GF12} for the compositionality results for bisimulation in \cite{LT05,LT09}.
Our aim is then to exploit the decomposition method of \cite{CGT16b} and the characterization of bisimilarity metric showed in this paper in order to systematically derive formats for bisimilarity metric, namely to obtain a logical characterization of the compositionality results for bisimilarity metric presented in \cite{GLT15,GT15,GLT16}.

Finally, we will apply our characterization approach to various behavioral metrics as \emph{convex bisimulation metric} \cite{SL95}, \emph{weak bisimulation metric} \cite{DJGP02} and \emph{trace metric} \cite{FL14}. 
We aim also to apply the approach to the notion of \emph{$\epsilon$-bisimulation} \cite{ABG04,ABdPGHW02,dPHW03}, for which a modal decomposition of formulae characterizing the compositional results in \cite{T08,T10} can be given.

\bibliographystyle{eptcs}
\bibliography{qapl16_bib}

\end{document}